\documentstyle[preprint,aps]{revtex}
\tightenlines
\begin{document}
\title{Autoregressive model of 1/f noise}
\author{B. Kaulakys}
\address{Institute of Theoretical Physics and Astronomy, A. Go\v stauto 12, LT-2600\\
Vilnius, Lithuania \\
and Department of Physics, Vilnius University, Saul\.etekio al. 9, LT-2040\\
Vilnius, Lithuania \\
\vspace{1cm} Physics Letters A Vol.257, no.1/2, p.37-42 (1999)
\\
\vspace{1cm}}
\date{Received 7 August 1998; received in revised form 15 February 1999;
accepted 28 April 1999}
\maketitle

\begin{abstract}
An analytically solvable model is proposed exhibiting 1/f spectrum in any
desirably wide range of frequency (but excluding the point $f=0$). The model
consists of pulses whose recurrence times obey an autoregressive process
with very small damping.

\noindent PACS: 05.40.+j, 02.50.-r, 72.70.+m

Keywords: 1/f noise, Brownian motion, time series, long-memory processes,
fluctuation phenomena
\end{abstract}

\pacs{05.40.+j, 02.50.-r, 72.70.+m}

{\bf 1. Introduction}

The puzzle of the origin and omnipresence of 1/f noise -- also known as
'flicker' or 'pink' noise -- is one of the oldest unsolved problem of the
contemporary physics. Since the first observation of flicker noise in the
currents of electron tubes more than 70 years ago by Johnson \cite{johnson25}
, fluctuations of signals and physical variables exhibiting behavior
characterized by a power spectral density $S\left( f\right) $ diverging at
low frequencies like $1/f^\delta $ ($\delta \simeq 1$) have been discovered
in large diversity of uncorrelated systems. We can mention here processes in
condensed matter, traffic flow, quasar emissions and music, biological,
evolution and artificial systems and even human cognition (see \cite
{hooge81,musha76,press78,koba82} and references herein).

1/f noise is an intermediate between the well understood white noise with no
correlation in time and the random walk (Brownian motion) noise with no
correlation between increments. The widespread occurrence of signals
exhibiting power spectral density with 1/f behavior suggests that a general
mathematical explanation of such an effect might exist. However, except for
some formal mathematical descriptions like ''fractional Brownian motion'' or
half-integral of a white noise signal \cite{mandelbr68} no generally
recognized physical explanation of the ubiquity of 1/f noise is still
proposed. Models of 1/f noise in some physical systems are usually
specialized (see \cite{hooge81,musha76,press78,koba82} and references
herein) and they do not explain the omnipresence of the processes with $
1/f^\delta $ spectrum \cite{icnf95,icnf97,upon97}.

Note also some mathematical algorithms and models of the generation of the
processes with 1/f noise \cite{jensen91,kumic94,sinha96}. These models can
not, as a rule, be solved analytically and they do not reveal the origin as
well as the necessary and sufficient conditions for the appearance of 1/f
type fluctuations.

History of the progress in different areas of physics points to the crucial
influence of simple models on the understanding of new phenomena. We note
here only the decisive influence of the Lorenz model as well as the logistic
and standard (Chirikov) maps for understanding of the deterministic chaos
and the quantum kicked rotor for revealing the quantum localization of
classical chaos.

It is the purpose of this letter to present and analyze the simplest
analytically solvable model of 1/f noise which can be relevant for the
understanding of the origin, main properties and parameter dependencies of
flicker noise. Our model is a result of the search for necessary and
sufficient conditions for the appearance of 1/f fluctuations in simple
systems affected by random external perturbations, originated from the
observation of a transition from chaotic to nonchaotic behavior in an
ensemble of randomly driven systems \cite{kaul95}, initiated in Ref. \cite
{kaul97} and further developed in Ref. \cite{kaul98}.

In the model there are analyzed currents or signals represented as sequences
of random but correlated pulses whose recurrence times (intervals between
transit times of pulses) obey an autoregressive process with small damping.
It is shown that for small average recurrence time and very small damping,
random increments of the recurrence times lead to 1/f behavior of the power
spectrum of the signal or current in wide range of frequency, however,
analytical at $f=0$.

{\bf 2. Model and solution}

Let us consider a point process when the intensity of some signal consisting
from a sequence of pulses (elementary events) or current of particles
through some Poincar\'e section $L_m$ may be expressed as
$$
I\left( t\right) =\sum_ka\delta \left( t-t_k\right) .\eqno{(1)}
$$
Here $\delta \left( t\right) $ is the Dirac delta function, $\left\{
t_k\right\} $ is a sequence of transit times $t_k$ at which the particles or
pulses cross the section $L_m$ and $a$ is a contribution to the signal or
current of one pulse or particle when it crosses the section $L_m$.

The power spectral density of the current (1) is
$$
S\left( f\right) =\lim \limits_{T\rightarrow \infty }\left\langle \frac{2a^2}
T\left| \sum_{k=k_{\min }}^{k_{\max }}e^{-i2\pi ft_k}\right| ^2\right\rangle
=\lim \limits_{T\rightarrow \infty }\left\langle \frac{2a^2}T
\sum_k\sum_{q=k_{\min }-k}^{k_{\max }-k}e^{i2\pi f\Delta \left( k;q\right)
}\right\rangle \eqno{(2)}
$$
where $\Delta \left( k;q\right) \equiv t_{k+q}-t_k$ is the difference of
transit times $t_{k+q}$ and $t_k$, $T$ is the whole observation time
interval, $k_{\min }$ and $k_{\max }$ are minimal and maximal values of
index $k$ in the interval of observation and the brackets $\left\langle
...\right\rangle $ denote the averaging over realizations of the process.

Let us analyze a process whose recurrence times $\tau _k=t_k-t_{k-1}$ follow
an autoregressive AR(1) process with offset $\bar \tau >0$, regression
coefficient $\alpha =1-\gamma $ and noise variance $\sigma ^2$. So, defining
by $\theta _k=\tau _k-\bar \tau $ a deviation of the recurrence time $\tau
_k $ from the steady state value $\bar \tau $, we have autoregressive
equations for the deviations $\theta _k$
$$
\theta _k=\alpha \theta _{k-1}+\sigma \varepsilon _k\eqno{(3a)}
$$
and for the recurrence times $\tau _k$
$$
\tau _k=\tau _{k-1}-\gamma \left( \tau _{k-1}-\bar \tau \right) +\sigma
\varepsilon _k.\eqno{(3b)}
$$
Here $\left\{ \varepsilon _k\right\} $ denotes a sequence of uncorrelated
normally distributed random variables with zero expectation and unit
variance (the white noise source) and $\sigma $ is the standard deviation of
the white noise. Note that the coefficient $\gamma $ has a sense of damping
(the relaxation rate of the recurrence times $\tau _k$ to the average value $
\bar \tau )$ and we will consider only processes with $\gamma \ll 1$ or $
\gamma =0$.

The recurrence equation for the transit times $t_k$ is
$$
t_k=t_{k-1}+\tau _k\eqno{(4)}
$$
where the recurrence times $\tau _k$ is defined by Eq. (3b).

The simplest interpretation of our model corresponds to one particle moving
along some orbit. The period of this motion fluctuates (due to external
random perturbations of the system's parameters) about some average value $
\bar \tau $. Some generalizations and extensions of the model and its
interpretation will be discussed below.

From Eqs. (3a) and (3b) follows explicit expressions for the deviation $
\theta _k$ and the recurrence time $\tau _k$,
$$
\theta _k=\theta _0\alpha ^k+\sigma \sum\limits_{j=1}^k\alpha
^{k-j}\varepsilon _j,\eqno{(5a)}
$$
$$
\tau _k=\bar \tau +\left( \tau _0-\bar \tau \right) \alpha ^k+\sigma
\sum_{j=1}^k\alpha ^{k-j}\varepsilon _j,\eqno{(5b)}
$$
with $\theta _0$ and $\tau _0$ being the initial values of the deviation and
of the recurrence time, respectively.

The variance of the recurrence time $\tau _k$ is
$$
\sigma _\tau ^2\left( k\right) \equiv \left\langle \tau _k^2\right\rangle
-\left\langle \tau _k\right\rangle ^2=\sigma ^2\left( 1-\alpha ^{2k}\right)
/\left( 1-\alpha ^2\right) \eqno{(6)}
$$

After some algebra we can easily obtain from Eqs. (4) and (5b) an explicit
expression for the transit times $t_k$,
$$
t_k=t_0+\sum\limits_{j=1}^k\tau _j=t_0+k\bar \tau +\frac \alpha \gamma
\left( \tau _0-\bar \tau \right) \left( 1-\alpha ^k\right) +\frac \sigma
\gamma \sum_{l=1}^k\left( 1-\alpha ^{k+1-l}\right) \varepsilon _l,\eqno{(7)}
$$
with $t_0$ being the initial time.

The power spectral density of the current according to Eq. (2) depends on
the statistics of the transit times difference $\Delta \left( k;q\right) $
which according to Eq. (7) is
$$
\Delta \left( k;q\right) \equiv t_{k+q}-t_k=q\bar \tau +\frac 1\gamma \left(
\tau _0-\bar \tau \right) \left( 1-\alpha ^q\right) \alpha ^{k+1}
$$
$$
+\frac \sigma \gamma \left[ \left( 1-\alpha ^q\right)
\sum\limits_{l=1}^k\alpha ^{k+1-l}\varepsilon
_l+\sum\limits_{l=k+1}^{k+q}\left( 1-\alpha ^{k+q+1-l}\right) \varepsilon
_l\right] ,\quad q\geq 0.\eqno{(8)}
$$

Random variable $\Delta \left( k;q\right) $ is a sum of two regular terms
and $k+q$ uncorrelated Gaussian random variables with zero expectations and
variances $\left( \frac \sigma \gamma \right) ^2\left( 1-\alpha ^q\right)
^2\alpha ^{2\left( k+1-l\right) }$ for $l=1,2,\ldots ,k$ and $\left( \frac
\sigma \gamma \right) ^2\left( 1-\alpha ^{k+q+1-l}\right) ^2$ for $
l=k+1,k+2,\ldots ,k+q$, respectively. Therefore, $\Delta \left( k;q\right) $
is a normally distributed random variable with the expectation
$$
\mu _\Delta \left( k;q\right) \equiv \left\langle \Delta \left( k;q\right)
\right\rangle =q\bar \tau +\frac 1\gamma \left( \tau _0-\bar \tau \right)
\left( 1-\alpha ^q\right) \alpha ^{k+1}\eqno{(9)}
$$
and the variance $\sigma _\Delta ^2\left( k;q\right) \equiv \left\langle
\Delta \left( k;q\right) ^2\right\rangle -\left\langle \Delta \left(
k;q\right) \right\rangle ^2$, which equals the sum of the variances of the
components,
$$
\sigma _\Delta ^2\left( k;q\right) =\left( \frac \sigma \gamma \right)
^2\left[ \left( 1-\alpha ^q\right) ^2\sum\limits_{j=1}^k\alpha
^{2j}+\sum\limits_{j=1}^q\left( 1-\alpha ^j\right) ^2\right]
$$
$$
=\left( \frac \sigma \gamma \right) ^2\left[ q-\frac{2\alpha \left( 1-\alpha
^q\right) }{1-\alpha ^2}-\frac{\alpha ^{2k+2}\left( 1-\alpha ^q\right) ^2}{
1-\alpha ^2}\right] ,\quad q\geq 0.\eqno{(10)}
$$
Here new summation indexes $j=k+1-l$ and $j=k+q+1-l$ of first and second
sums in (8), respectively, have been introduced.

Note that from the definition of the difference of transit times it follows
the symmetry relations
$$
\Delta \left( k;-q\right) =-\Delta \left( k-q;q\right) ,\quad \mu \left(
k;-q\right) =-\mu \left( k-q;q\right) ,\quad \sigma _\Delta ^2\left(
k;-q\right) =\sigma _\Delta ^2\left( k-q;q\right) .\eqno{(11)}
$$

At $k\gg \gamma ^{-1}$ or after averaging over $\tau _0$ from the
distribution with the expectation $\bar \tau $ and variance $\sigma _\tau
^2\equiv \sigma _\tau ^2\left( \infty \right) =\sigma ^2/\left( 1-\alpha
^2\right) \simeq \sigma ^2/2\gamma $ according to Eq. (6), expressions (7)
and (8) generate a stationary time series. The expectation and the variance
of the difference $\Delta \left( k;q\right) $ of transit times of the
stationary time series are
$$
\mu _\Delta \left( q\right) \equiv \left\langle \mu _\Delta \left( \infty
;q\right) \right\rangle =q\bar \tau ,\eqno{(12)}
$$
$$
\sigma _\Delta ^2\left( q\right) \equiv \sigma _\Delta ^2\left( \infty
;q\right) =\left( \frac \sigma \gamma \right) ^2\left[ q-\frac{2\alpha
\left( 1-\alpha ^q\right) }{1-\alpha ^2}\right] .\eqno{(13)}
$$
The power spectral density of the current according to Eq. (2) is
$$
S\left( f\right) =\lim \limits_{T\rightarrow \infty }\frac{2a^2}T
\sum\limits_{k,q}\left\langle e^{i2\pi f\Delta \left( k;q\right)
}\right\rangle =\lim \limits_{T\rightarrow \infty }\frac{2a^2}T
\sum\limits_{k,q}\chi _{\Delta \left( k;q\right) }\left( 2\pi f\right)
\eqno{(14)}
$$
where $\chi _{\Delta \left( k;q\right) }\left( 2\pi f\right) $ is the
characteristic function of the distribution of the transit times difference $
\Delta \left( k;q\right) $. For the normal distribution of $\Delta \left(
k;q\right) $ the characteristic function takes the form $\chi _{\Delta
\left( k;q\right) }\left( 2\pi f\right) =\exp \left[ i2\pi f\mu _\Delta
\left( k;q\right) -2\pi ^2f^2\sigma _\Delta ^2\left( k;q\right) \right] $
and the power spectral density equals
$$
S\left( f\right) =\lim \limits_{T\rightarrow \infty }\frac{2a^2}T
\sum_{k,q}\exp \left[ i2\pi f\mu _\Delta \left( k;q\right) -2\pi ^2f^2\sigma
_\Delta ^2\left( k;q\right) \right] .\eqno{(15)}
$$

For $q\ll \gamma ^{-1}$ we have from Eqs. (9) and (10) expansions of the
expectation $\mu _\Delta \left( k;q\right) $ and of the variance $\sigma
_\Delta ^2\left( k;q\right) $ in powers of $\gamma q\ll 1$
$$
\mu _\Delta \left( k;q\right) =\tau \left( k\right) q,\quad \tau \left(
k\right) =\bar \tau +\left( \tau _0-\bar \tau \right) \alpha ^{k+1},
\eqno{(16)}
$$
$$
\sigma _\Delta ^2\left( k;q\right) =\frac{\sigma ^2}2\left\{ \left[ \frac{
2\left( 1-\alpha ^{2k}\right) }{1-\alpha ^2}+\alpha ^{2k}\right] q^2+\left(
\alpha ^{2k}-\frac 13\right) q^3+\frac 13q\right\} .\eqno{(17)}
$$
The leading term of the expansion (17) may be written as
$$
\sigma _\Delta ^2\left( k;q\right) =\sigma _\tau ^2\left( k\right) q^2,\quad
\gamma q\ll 1\eqno{(18)}
$$
where the variance $\sigma _\tau ^2\left( k\right) $ of the recurrence time $
\tau _k$ is defined by Eq. (6). Therefore, Eq. (15) takes the form
$$
S\left( f\right) =\lim \limits_{T\rightarrow \infty }\frac{2a^2}T
\sum_{k,q}\exp \left[ i2\pi f\tau \left( k\right) q-2\pi ^2f^2\sigma _\tau
^2\left( k\right) q^2\right] .\eqno{(19)}
$$
Here the mean recurrence time $\tau \left( k\right) $ is defined in Eq. (16).

Eq. (19) is valid if $2\pi ^2f^2\sigma _\tau ^2\left( k\right) q^2\mid
_{q=\gamma ^{-1}}\gg 1$, i.e., for $f>f_1=\gamma /\pi \sigma _\tau \left(
k\right) $. When $f\ll f_\tau =\left[ 2\pi \tau \left( k\right) \right]
^{-1} $ and $f<f_2=\left[ \pi \sigma _\tau \left( k\right) \right] ^{-1}$ we
can replace the summation over $q$ in Eq. (19) by the integration. The
integration yields to the 1/f spectrum
$$
S\left( f\right) =\frac{a^2}f\sqrt{\frac 2\pi }\lim \limits_{T\rightarrow
\infty }\frac 1T\sum_k\frac 1{\sigma _\tau \left( k\right) }\exp \left[ -
\frac{\tau \left( k\right) ^2}{2\sigma _\tau ^2\left( k\right) }\right]
,\quad f_1<f<\min \left\{ f_2,f_\tau \right\} .\eqno{(20)}
$$

Eq. (20) is valid for the stationary as well as for the non-stationary
process with slowly changing of the expectation and variance of the
recurrence time.

At $k\gg \gamma ^{-1}$ or after averaging over $\tau _0$ from the
distribution with the expectation $\bar \tau $ and variance $\sigma _\tau
^2=\sigma ^2/\left( 1-\alpha ^2\right) \simeq \sigma ^2/2\gamma $ we have
the stationary process: the expectation and the variance of the recurrence
time do not depend on the parameter $k$, i.e., $\tau \left( k\right) =\bar
\tau $ and $\sigma _\tau \left( k\right) =\sigma _\tau $. For the stationary
process Eq. (20) takes the form
$$
S\left( f\right) =\bar I^2\frac{\alpha _H}f.\eqno{(21)}
$$
Here $\bar I=\lim \limits_{T\rightarrow \infty }a\left( k_{\max }-k_{\min
}+1\right) /T=a/\bar \tau $ is the average current and $\alpha _H$ is a
dimensionless constant (the Hooge parameter)
$$
\alpha _H=\frac 2{\sqrt{\pi }}Ke^{-K^2},\quad K=\frac{\bar \tau }{\sqrt{2}
\sigma _\tau }.\eqno{(22)}
$$

Therefore, the power of 1/f noise except of the squared average current
strongly depends on the ratio of the average recurrence time to the standard
deviation of the recurrence time.

{\bf 3. Discussion and generalizations }

The point process containing only one relaxation time $\gamma ^{-1}$ can for
sufficiently small damping $\gamma $ and average recurrence time $\bar \tau
\ll \sigma /\sqrt{\gamma }$ (with $\sigma $ being the standard deviation of
the white noise source) produce an exact 1/f-like spectrum in wide range of
frequency $\left( f_1,f_2\right) $, with $f_2/f_1\simeq \gamma ^{-1}$.
Furthermore, due to the contribution to the transit times $t_k$ of the large
number of the random variables $\varepsilon _l$ ($l=1,2,...k$), our model
represents a 'long-memory' random process. As a result of the nonzero
relaxation rate ($\gamma \neq 0$) and, consequently, due to the finite
variance, $\sigma _\tau ^2=\sigma ^2/2\gamma $, of the recurrence time the
model is free from the unphysical divergency of the spectrum at $
f\rightarrow 0$. So, using an expansion of expression (13) at $\gamma q\gg 1$
, $\sigma _\Delta (q)=\left( \sigma /\gamma \right) ^2q$, we obtain from Eq.
(15) the spectrum density $S(f)=\bar I^2\left( 2\sigma ^2/\bar \tau \gamma
^2\right) $ for $f\ll \min \left\{ f_1,f_0=\bar \tau \gamma ^2/\pi \sigma
^2\right\} $. This is in agreement with the statement \cite{Schick74} that
the power spectrum of any pulse sequence is white at low enough frequencies.

This simple, consistent and exactly solvable model can easily be generalized
in different directions: for large number of particles moving in similar
orbits with coherent (identical for all particles) or independent
(uncorrelated for different particles) fluctuations of the periods, for
non-Gaussian or continuous perturbations of the systems' parameters, for
nonlinear relaxation and for spatially extended systems. So, when an
ensemble of $N$ particles moves on closed orbits and the period of each
particle fluctuates independently (due to the perturbations by uncorrelated
sequences of random variables $\{\varepsilon _k^\upsilon \}$, different for
each particle $\upsilon $) the power spectral density of the collective
current $I$ of all particles can be calculated by the above method too and
is expressed as the Hooge formula \cite{hooge81}
$$
S\left( f\right) =\bar I^2\frac{\alpha _H}{Nf}.\eqno{(23)}
$$

The model may be used for evaluation of the power spectral density of the
non-stationary process as well. So, at $k\ll \gamma ^{-1}$ or for $\gamma =0$
we have a process with the constant averaged recurrence time $\tau \left(
k\right) =\tau _0$, from Eq. (16), and linearly increasing variance of the
recurrence time $\sigma _\tau ^2\left( k\right) =\sigma ^2k$, according to
Eq. (6), i.e., a process similar to the Brownian motion without relaxation.
For $k\gg \left| q\right| $ it is valid expansion (18) and the power
spectral density of such process for finite observation time interval $T$
may be evaluated according to Eq. (20)
$$
S\left( f\right) =\frac 1f\sqrt{\frac 2\pi }\frac{a^2}{\sigma T}
\sum_{k=k_{\min }}^{k_{\max }}\frac 1{\sqrt{k}}\exp \left[ -\frac{\tau _0^2}{
2\sigma ^2k}\right] ,\quad f_1\ll f<f_2.\eqno{(24)}
$$
Here
$$
f_1=\left( \sigma k_{\max }\sqrt{k_{\min }}\right) ^{-1},\quad f_2=\left(
\pi \sigma \sqrt{k_{\max }}\right) ^{-1}.\eqno{(25)}
$$

The process with random increments of the recurrence time and without
relaxation is, however, very unstable and strongly depending on the
realization. Really, the averaged number, $\left\langle k_{\max }-k_{\min
}\right\rangle $, of transition times $t_k$ for the given observation time
interval $T$ is $\left\langle k_{\max }-k_{\min }\right\rangle =T/\tau _0$
but the standard deviation $\sigma _T$ of the time interval for given $
k_{\min }$ and $k_{\max }$ according to Eq. (10) equals $\sigma _T=\sigma
\left( k_{\max }-k_{\min }\right) \sqrt{\left( k_{\max }+2k_{\min }\right) /3
}$. Therefore, for $\left( k_{\max }+2k_{\min }\right) >3\left( \tau
_0/\sigma \right) ^2$ the standard deviation exceeds the expectation value $
\left\langle T\right\rangle =\left( k_{\max }-k_{\min }\right) \tau _0$ of
the time interval. The power spectrum of any realization is, however, of 1/f
type for large frequency interval $\left( f_1,f_2\right) $, with $
f_2/f_1\simeq \sqrt{k_{\max }k_{\min }}$.

It should be noticed in conclusion that in many cases the intensity of
signals or currents can be expressed in the form (1). This expression
represent exactly the flow of identical point objects. More generally, in
Eq. (1) instead of the Dirac delta function one should introduce time
dependent pulse amplitudes $A_k\left( t-t_k\right) $. However, the low
frequency power spectral density depends weakly on the shapes of the pulses
\cite{Schick74}, while fluctuations of the pulses amplitudes result, as a
rule, in white or Lorentzian, but not 1/f, noise. The model (2)--(5) in such
cases represents fluctuations of the averaged period $\tau _k$ between the
subsequent transition times of the pulses. Therefore, the model may be
easily generalized and applied for the explanation of 1/f noise in different
systems. Furthermore, it reveals the possible origin of 1/f noise, i.e.
random increments of the time intervals between the pulses or elementary
events.

Summarizing, a simple analytically solvable model of 1/f noise is presented
and analyzed. The model reveals main features and parameter dependences of
the power spectral density of the noise. The model and its generalizations
may essentially influence the understanding of the origin and main
properties of the flicker noise.

{\bf Acknowledgements}

The author acknowledges stimulating discussions with Prof. R. Katilius,
Prof. A. Matulionis, Dr. A. Bastys and Dr. T. Me\v skauskas. The research of
this publication was made possible in part by support of the Alexander von
Humboldt Foundation and Lithuanian State Science and Studies Foundation.

\end{document}